\documentclass[twocolumn,showpacs,preprintnumbers,amsmath,amssymb,pra]{revtex4}
\usepackage{graphicx}
\usepackage{dcolumn}
\usepackage{bm}
\def\rv#1{{#1 \rangle}}
\def\rrv#1{{#1 \rangle\rangle}}
\def\lv#1{{\langle #1}}
\def\llv#1{{\langle\langle #1}}
\def\so#1{{\cal #1}}
\begin{document}

\title{Completely-Positive Non-Markovian Decoherence}

\author{Jinhyoung Lee}

\email{hyoung@hanyang.ac.kr}

\affiliation{Quantum Photonic Science Research Center and Department of
  Physics, Hanyang University, Seoul 133-791, Korea}

\affiliation{School of Mathematics and Physics, Queen's University,
  Belfast BT7 1NN, United Kingdom}

\author{Inbo Kim}

\affiliation{Institute of Quantum Information Processing and Systems,
  University of Seoul, Seoul, Korea}

\author{Helen McAneney}

\affiliation{School of Mathematics and Physics, Queen's University,
  Belfast BT7 1NN, United Kingdom}

\author{M. S. Kim}

\affiliation{School of Mathematics and Physics, Queen's University,
  Belfast BT7 1NN, United Kingdom}

\author{Doyeol Ahn}

\affiliation{Institute of Quantum Information Processing and Systems,
  University of Seoul, Seoul, Korea}

\date{\today}

\begin{abstract}
  We propose an effective Hamiltonian approach to investigate
  decoherence of a quantum system in a non-Markovian reservoir,
  naturally imposing the complete positivity on the reduced dynamics of
  the system. The formalism is based on the notion of an effective
  reservoir, {\em i.e.}, certain collective degrees of freedom in the
  reservoir that are responsible for the decoherence.  As examples for
  completely positive decoherence, we present three typical decoherence
  processes for a qubit such as dephasing, depolarizing, and
  amplitude-damping.  The effects of the non-Markovian decoherence are
  compared to the Markovian decoherence.
\end{abstract}


\maketitle

\section{Introduction}

An open quantum system has been extensively studied for potential
applications to quantum computation and quantum information processing
\cite{Jozsa94,Schumacher96,Lidar98,Lee00}. Through the interaction with
a reservoir a quantum system loses its coherent information. This is a
so-called {\em decoherence process} \cite{Zurek03}. Decoherence has been
regarded as a critical obstacle against quantum information processing.
However, as the notion of ``reservoir engineering'' has been suggested
for laser cooled trapped ions \cite{Cirac96}, the studies on the
decoherence have moved to control it in laboratories instead of
suffering from it \cite{Myatt00}.  Many of these studies have been based
on Markovian reservoirs \cite{Gardiner92}. A Markovian reservoir is
characterized by two essential properties: a) {\em weak} coupling with a
system by which Born approximation is validated and b) {\em rapid}
relaxation time such that the information of the system is diffused over
the reservoir in a rather short time compared to the time scale in which
the system changes. In the perspective of engineering a reservoir, it is
desirable to study the decoherence caused through a non-Markovian
reservoir. It is strongly believed that a solid-state realization of
quantum information processing may be extremely useful \cite{Kane98}.
The decoherence phenomena for solid state systems including a photonic
band gap material and a quantum dot cannot always be understood under
the Markovian assumption \cite{John97}.  The evolution of a single-mode
atomic cavity in a non-Markovian reservoir has been studied for an atom
laser \cite{Hope00}.

A non-Markovian reservoir has ``memory effects'' such that it
preserves the coherent information of the system within its
relaxation time. In order to analyze how many operations can be
performed by preserving the coherent information of the system,
one may consider three types of characteristic times: operation
time $\tau_o$ for a single quantum operation on the system,
decoherence time $\tau_d$ due to the interaction between the
system and the reservoir, and the relaxation time $\tau_r$ within
the reservoir.  When a quantum system decoheres in a Markovian
reservoir with $\tau_r \ll \tau_d$, the time correlation of
fluctuations in the reservoir is neglected. However, in a
non-Markovian reservoir with $\tau_r$ comparable to $\tau_d$, it
is expected that the time correlation between fluctuations during
the time interval $\Delta \tau \lesssim \tau_r$ raises a
correlated influence on the system. Once perturbed by the system,
the reservoir may memorize a part of the system's information
during $\tau_r$ \cite{Ahn97,Knezevic03}. This memorized information will be
fed back to the system at another perturbation within $\tau_r$. It
is expected that the correlated influence significantly suppresses
the decoherence of the system \cite{Ahn02}. An extremal example
for the memory effects by a non-Markovian reservoir is that two
atoms, which are prepared initially in a product states, come to
be in an entangled state through the interaction with a common
thermal reservoir \cite{Lee02,Braun02}. In Ref.~\cite{Lee02}, the
reservoir consists of a single-mode field with no relaxation,
$\tau_r \rightarrow \infty$. It is desirable to study decoherence
processes by bridging the gap between the two limiting cases of  $
\tau_r/\tau_d \rightarrow 0$ and $\tau_r/\tau_d \rightarrow
\infty$.

Dynamics of an open system may be obtained from a quantum Liouville
equation of the total system that consists of the system and the
reservoir \cite{Gardiner92, Zurek03}. The reduced density operator of
the system needs to be positive at all times. The positivity does not
need to be altered by the presence of any other system, in other words,
the reduced density operator of the system needs to be completely
positive if the system and the reservoir are initially uncorrelated
\cite{Stinespring55,Kraus71,Pechukas94}. The complete positivity was
intensively discussed for a Markovian reservoir \cite{Lindblad76}. For a
non-Markovian reservoir, it was discussed for a master equation derived
by using an operator sum representation with coarse-graining
approximation \cite{Bacon99}. A random telegraph signal model was
studied for a system in a non-Markovian reservoir \cite{Daffer03}. However the
complete positivity of the system density operator in a non-Markovian
reservoir is still an important problem to be unraveled.

In this paper, we propose an effective Hamiltonian approach to
investigate the decoherence of a quantum system in a non-Markovian
reservoir.  Instead of deriving a non-Markovian master equation, the
formalism is based on the notion of an effective reservoir, {\em i.e.},
certain collective degrees of freedom in the reservoir that is
responsible for the decoherence.  We show that {\em the reduced dynamics
  of the open system is completely positive} and the complete positivity
is naturally imposed from the effective Hamiltonian approach. As
examples for completely positive decoherence, we present three typical
decoherence processes for a qubit such as dephasing, depolarizing, and
amplitude-damping. The effects of the non-Markovian decoherence are
compared to the Markovian decoherence.

\section{Effective environment}

\label{sec:ee}

Dynamics of an open quantum system has commonly been investigated by a
master equation for the density operator or a Fokker-Planck equation for
the quasi-probability function such as a Wigner function
\cite{Gardiner92}.  It is however difficult to derive a completely
positive master equation for a system interacting with a general
non-Markovian reservoir.
We suggest an effective Hamiltonian approach by introducing the notion
of ``effective variables'' in the environment. It will be shown that the
large (normally infinite) number of environmental degrees of freedom,
which we call ``environmental variables'', is reductive into a small
number of {\em effective environmental variables} in the sense that both
cases result in the same master (or Fokker-Planck) equation for the
system.  For instance, a Markovian thermal environment is reductive into
a collective single-mode boson field in a thermal state
\cite{McAneney03}.


The property of being reductive into the small effective
environmental variables, that leads to the correct and equivalent
description of the reduced dynamics for a system, is predicted by
two observations.  First, at a given time $\tau$, the density
operator $\hat{\rho}$ of the system $S$ can always be purified to
a pure state $|\psi\rangle$ of a larger composite system
consisting of the system $S$ and an ancillary system $R$, such
that $\hat{\rho}$ is obtained by tracing the pure state over the
ancillary system, {\em i.e.}, $\hat{\rho} = \mbox{Tr}_R
|\psi\rangle_{SR}\langle\psi|$. For the purification, an ancillary
system is required to have its Hilbert space larger than or equal
to that of the system \cite{Jozsa94}: For instance, a qubit may
suffice for the ancillary system if the system is a qubit. The
composite system $S+R$ provides all physical descriptions relating
to the system and the ancillary system is called effective
variable(s), which is in effect certain collective degrees of
freedom, of the environment at the given time $\tau$.  Secondly,
the effective variables may be dynamic over the whole
environmental variables as the interaction time $\tau$ passes.  On
the other hand, the dynamics of the effective variables can be
absorbed by time-dependent coupling constants while keeping the
effective variables stationary. The set of the time-dependent
coupling constants and the effective variables contains all the
information of the environment that governs the decoherence of the
system.  We call this set an effective environment.

Suppose $\hat{H}_{int}(\tau)$ is an interaction Hamiltonian
between a system $S$ and an effective environment $R$. We write the
interaction Hamiltonian $\hat{H}_{int}(\tau)$, in the interaction
picture, in the form of
\begin{eqnarray}
  \label{eq:eeiih}
  \hat{H}_{int}(\tau) = \sum_{\alpha} \hat{S}_\alpha \otimes \hat{D}_\alpha(\tau)
\end{eqnarray}
where $\hat{S}_\alpha$ is a Hermitian operator for the system $S$ and
$\hat{D}_\alpha(\tau) = \sum_\beta \lambda_{\alpha\beta}(\tau)
\hat{R}_\beta$ with Hermitian operator $\hat{R}_\beta$ is for the
effective environment $R$. The unit of $\hbar = 1$ is used throughout
the paper. Here, the time-dependent coupling constant
$\lambda_{\alpha\beta}(\tau)$ is a Hermitian matrix.  The composite
system $S+R$ is assumed initially in a product state,
\begin{eqnarray}
  \label{eq:ist}
  \hat{\rho}_T(0) = \hat{\rho}_S(0)\otimes\hat{\rho}_R(0).
\end{eqnarray}
The effective environment $R$ consists of collective degrees of freedom
in the environment $E$ \cite{McAneney03}. Its initial state
$\hat{\rho}_R(0)$ is not necessarily an thermal state even when the real
environment is thermal. The Hilbert space of the effective environment
and the time-dependent interaction Hamiltonian $\hat{H}_{int}(\tau)$ are
determined such that they result in the correct reduced dynamics of the
system. They may be determined from first principle calculation such as
the time-convolutionless projection-operator method \cite{Ahn02}. The
use of the time-convolutionless projection-operator method is presented
in Sec.~III.

Once the effective environment is determined, one may consider and solve
the quantum Liouville equation for the composite system,
\begin{eqnarray}
  \label{eq:qlet}
  \frac{d}{d\tau} \hat{\rho}_T(\tau) = -i
  [\hat{H}_{int}(\tau),\hat{\rho}_T(\tau)].
\end{eqnarray}

The evolution operator is now given as
\begin{eqnarray}
  \label{eq:fseo}
  \hat{U}_{int}(\tau) &=& {\mathbf T} \exp \left( -i \int_0^\tau d\tau'
    \hat{H}_{int}(\tau') \right)
\end{eqnarray}
where ${\mathbf T}$ is the time-ordering operator. In most cases,
it is convenient to let $\lambda_{\alpha\beta}(\tau) =
\lambda(\tau) g_{\alpha\beta}$ or $\hat{D}_\alpha(\tau) =
\lambda(\tau) \hat{D}_\alpha$ so that $\hat{U}_{int}(\tau) = \exp
[-i \Lambda(\tau) \sum_{\alpha}
\hat{S}_\alpha\otimes\hat{D}_\alpha]$ where
$\Lambda(\tau)=\int_0^\tau d\tau' \lambda(\tau')$. This simpler
form of $\hat{U}_{int}(\tau)$ has advantages for later discussions
from the calculational point of view.

The total system is described by the density operator
\begin{eqnarray}
  \label{eq:tdo}
  \hat{\rho}_T(\tau) = \hat{U}_{int}(\tau) \hat{\rho}_T(0) \hat{U}_{int}^\dag(\tau),
\end{eqnarray}
and the reduced density operator of $S$ is given by partially tracing
the total density operator over $R$,
\begin{eqnarray}
  \label{eq:rdos}
  \hat{\rho}_S(\tau) = \mbox{Tr}_R \hat{\rho}_T(\tau).
\end{eqnarray}
Letting $\{|n\rangle\}$ be the orthonormal basis set of $R$ that
diagonalizes $\hat{\rho}_R(0) = \sum_n p_n |n \rangle_R\langle n|$, the
reduced dynamics of the system is described by a Kraus representation
(or operator sum representation) \cite{Kraus71}:
\begin{eqnarray}
  \label{eq:osrrdo}
  \hat{\rho}_S(\tau) = {\cal S}(\tau) \hat{\rho}_S(0) = \sum_{n,m}
  \hat{K}_{nm}(\tau) \hat{\rho}_S(0) \hat{K}_{nm}^\dag(\tau)
\end{eqnarray}
where $\hat{K}_{nm}(\tau) = {}_R\langle n| \hat{U}_{int}(\tau) |m
\rangle_R\sqrt{p_{m}}$. Note that the superoperator ${\cal S}$ is a
linear operator in the Hilbert-Schmidt space of density operators.
Calligraphic letters are used for superoperators on the Hilbert-Schmidt
space throughout the paper. It is remarkable that the existence of the
Kraus representation $\{\hat{K}_{nm}(\tau)\}$ directly implies the
complete positivity of the evolution superoperator $\so{S}(\tau)$
\cite{Kraus71}.  This fact was already guaranteed since the Hamiltonian
formalism was adopted in the present approach of effective environment.
This approach is applicable to the system decohering in a non-Markovian
as well as a Markovian environment.

The present approach of effective environment has several advantages in
describing the reduced dynamics of a system. Firstly, it provides the
correct and equivalent description, similar to the master equation.
Secondly and most importantly, it guarantees the complete positivity in
a general environment like a non-Markovian environment. Thirdly, it
enables the analysis of the structure for effective variables, in a
given environment, that are directly responsible for the decoherence.
This analysis will be given in Sec.~\ref{sec:qubit}. Lastly, one may
simulate the decoherence in an experiment by introducing such an
effective environment and controlling the coupling to the given system.

\section{Determination of effective environment}

\label{sec:dee}

In the previous section we showed that the the effective
environment approach provides the description of the reduced
dynamics of a quantum system, which decoheres through an
environment, provided the Hilbert space of the effective
environment, its initial state $\hat{\rho}_R(0)$ and the
time-dependent interaction Hamiltonian $\hat{H}_{int}(\tau)$ are
determined. In this section we suggest a scheme which determines
them by a first principle theory, in particular, the
time-convolutionless projection-operator method
\cite{Ahn94,Ahn97,Ahn00} which has been employed to study
non-Markovian environments for quantum information processing
\cite{Ahn02}.

\subsection{Time-convolutionless projection-operator scheme}

Consider the total system of a quantum system and an environment. The
total system is assumed to have the Hamiltonian,
\begin{eqnarray}
{\hat H}_T={\hat H}_S+{\hat H}_E + {\hat H}_{int},
\end{eqnarray}
where ${\hat H}_S$, ${\hat H}_E$, and ${\hat H}_{int}$ are the
Hamiltonians for the system, the environment, and their interaction
respectively. It is straightforward to generalize that ${\hat H}_S$ may
contain time-dependent external fields to control a quantum operation on
the system.  For a typical form of the interaction Hamiltonian, we
consider a Caldeira-Leggett-type model~\cite{Caldeira85} given by
\begin{eqnarray}
  \label{eq:cltm}
  {\hat H}_{int} = \sum_{\alpha} {\hat S}_\alpha \otimes {\hat E}_\alpha
\end{eqnarray}
where ${\hat S}_\alpha$ is a Hermitian operator acting on the system and
${\hat E}_\alpha = \sum_k (g_{\alpha k} {\hat e}^\dagger_k + g_{\alpha
  k}^* {\hat e}_k)$ is a fluctuating boson field due to perturbation
arising from the system. Here, $\hat{e}_k$ ($\hat{e}^\dag_k$) is
an annihilation (creation) operator for a boson mode $k$. The
unperturbed boson fields of the environment are governed by ${\hat
H}_E = \sum_{k} \omega_k {\hat e}^{\dagger}_k {\hat e}_k $.  The
set of the operators $\{{\hat E}_\alpha\}$ describes various
decoherence processes.

The quantum Liouville equation for the density operator ${\tilde
  \rho}_T$ of the total system is given by
\begin{eqnarray}
  \label{eq:qle}
  \frac{d}{d\tau}{\tilde \rho}_T(\tau) = -i[{\hat H}_T,{\tilde \rho}_T(\tau)] =
  -i \so{L}_T {\tilde \rho}_T(\tau),
\end{eqnarray}
where ${\cal L}_T={\cal L}_S + {\cal L}_E + {\cal L}_{int}$ is the
Liouville operator. Here, the symbol tilde is given to indicate a
density operator in the Sch\"odinger picture. Each Liouville operator
${\cal L}$ is a superoperator on the Hilbert-Schmidt space of density
operators. The Liouville operators are in one-to-one correspondence to
the Hamiltonians of the same subscriptions. Before the system starts to
decohere, the environment is assumed initially in a thermal state
\begin{equation}
  \label{eq:trs}
  \tilde{\rho}_E = \frac{1}{Z}\exp\left(- \hat{H}_E/k_B T \right)
\end{equation}
where $Z=\mbox{Tr} \exp (- \hat{H}_E/k_B T)$, $T$ is the temperature and
$k_B$ the Boltzmann constant.  The assumption may be released to all the
time-independent states that commute with the non-interacting
Hamiltonian of the environment, {\em i.e.}, ${\cal L}_E
\tilde{\rho}_E=0$.

In order to derive and solve a reduced equation for the system alone, we
employ a projection-operator method \cite{Zwanzig60}. Time-independent
projection operators ${\cal P}$ and ${\cal Q}$ are defined as
\begin{eqnarray}
{\cal P} {\hat X}= {\tilde \rho}_E {\rm Tr}_E({\hat X}),~~
{\cal Q}=1-{\cal P},
\end{eqnarray}
for a dynamical variable ${\hat X}$ on the total system. Here ${\rm
  Tr}_E$ indicates a partial trace over the environment. The reduced
density operator of the system is given by ${\tilde \rho}_S(\tau) = {\rm
  Tr}_E {\tilde \rho}_T(\tau)= {\rm Tr}_E [{\cal P} {\tilde
  \rho}_T(\tau)$].

The quantum Liouville equation (\ref{eq:qle}) can be decomposed into two
coupled equations for ${\cal P}\tilde{\rho}_T$ and ${\cal
  Q}\tilde{\rho}_T$ respectively by applying the projection operators.
For the total system decoupled at $\tau=0$, the solution for ${\cal
  Q}\tilde{\rho}_T$ is substituted for the equation of ${\cal
  P}\tilde{\rho}_T$ with an ansatz
\begin{eqnarray}
  \label{eq:ans}
  {\cal P}{\cal L}_{int}{\cal P}=0.
\end{eqnarray}
The ansatz is introduced to ignore the renormalization of the
unperturbed energy \cite{Zwanzig60} (otherwise it raises the Lamb shift
\cite{Lamb}). Represented in the interaction picture, the
time-convolutionless master equation for the system is given
\cite{Ahn97} by
\begin{equation}
  \label{eq:seq}
  \frac{d}{d\tau}{\hat \rho}_S(\tau) = {\cal C}(\tau) {\hat \rho}_S(\tau),
\end{equation}
where ${\hat \rho}_S(\tau) = \exp(i \tau {\cal L}_S)
\tilde{\rho}_S(\tau)$ is the reduced density operator in the interaction
picture and ${\cal C}(\tau)$ is the generalized collision operator.

Let us consider a weak-coupling approximation (or a short-time limit) up
to the second-order $({\hat H}_{int})^2$. The collision operator ${\cal
  C}(\tau)$ can now be written as
\begin{eqnarray}
  \label{eq:socol2}
  {\cal C}(\tau) {\hat \rho}_S(\tau) &=& \sum_{\alpha\beta} \int_0^\tau d\tau'
  \Big\{\tilde{\chi}_{\alpha\beta}(\tau-\tau') [{\hat S}_\beta(\tau'){\hat \rho}_S(\tau),
  {\hat S}_\alpha(\tau)] \nonumber \\
  &+& \tilde{\chi}_{\alpha\beta}(\tau'-\tau)[{\hat S}_\beta(\tau),
  {\hat \rho}_S(\tau){\hat S}_\alpha(\tau')] \Big\}
\end{eqnarray}
where ${\hat S}_\alpha(\tau)=\exp(i\tau {\cal L}_S)\hat{S}_\alpha$ and
\begin{eqnarray}
   \label{eq:chif}
   \tilde{\chi}_{\alpha\beta}(\tau) = {\rm Tr}_E\left[{\hat E}_{\alpha}(\tau){\hat
   E}_{\beta}{\tilde \rho}_E\right] = \tilde{\chi}_{\beta\alpha}^*(-\tau)
\end{eqnarray}
with ${\hat E}_{\alpha}(\tau)=\exp(i\tau{\cal L}_E) {\hat E}_{\alpha}$.
The function $\tilde{\chi}_{\alpha\beta}(\tau)$ characterizes all
properties of the environment for the decoherence of the system. In
particular, it describes the time correlation between the quantum
fluctuations $\hat{E}_\alpha(\tau)$ and $\hat{E}_\beta(0)$ perturbed by
the system at the respective times. The time convolutionless form of
Eq.~(\ref{eq:seq}) is one of the crucial advantages of using the
projection-operator scheme while a time-convolution equation would be
derived by a simple perturbation theory \cite{Barnett97}. It was shown
\cite{Ahn02} that the time-convolutionless equation (\ref{eq:seq})
becomes the Lindblad master equation in the Markov approximation.

\subsection{Connection to effective environment approach}

In order to make a connection to the effective environment scheme
presented in Sec.~\ref{sec:ee}, Eq.~(\ref{eq:socol2}) will be further
analyzed. In Eq.~(\ref{eq:socol2}) the system operators depend on the
evolution time.  The time-dependent operators can be expanded by a
complete set of Hermitian operators, $C$, in the Hilbert-Schmidt space
where the set $C$ is chosen to include the operators $\hat{S}_\alpha$ of
the system:
\begin{eqnarray}
  \label{eq:ebcho}
  \hat{S}_\alpha(\tau) = \sum_{\beta=1}^{d^2-1} c_{\alpha\beta}(\tau)
  \hat{S}_\beta.
\end{eqnarray}
where $d$ is the dimension of the systems Hilbert space.  The
coefficient $c_{\alpha\beta}(\tau)$ can be obtained by solving the
Heisenberg equation,
\begin{eqnarray}
  \label{eq:hefo}
  \frac{d}{d\tau}\hat{S}_\alpha(\tau) = i[\hat{H}_S,\hat{S}_\alpha(\tau)].
\end{eqnarray}
Eqs.~(\ref{eq:ebcho}) and (\ref{eq:hefo}) give the following equation,
which is equivalent to Eq.~(\ref{eq:socol2}),
\begin{eqnarray}
  \label{eq:socol2p}
  {\cal C}(\tau) {\hat \rho}_S(\tau) &=& \sum_{\alpha\beta=1}^{d^2-1}
  \Big\{\gamma_{\alpha\beta}(\tau) [{\hat S}_\beta{\hat \rho}_S(\tau),
  {\hat S}_\alpha] \nonumber \\
  &&+ \gamma_{\beta\alpha}^*(\tau)[{\hat S}_\beta,
  {\hat \rho}_S(\tau){\hat S}_\alpha] \Big\}
\end{eqnarray}
where
\begin{eqnarray}
  \label{eq:chipom}
  \gamma_{\alpha\beta}(\tau) &=& \int^\tau_0 d\tau'
  \chi_{\alpha\beta}(\tau,\tau') \\
  \gamma_{\beta\alpha}^*(\tau) &=& \int^\tau_0 d\tau'
  \chi_{\alpha\beta}(\tau',\tau) \\
  \chi_{\alpha\beta}(\tau,\tau') &=& \sum_{\gamma\delta}
  c_{\gamma\alpha}(\tau) \tilde{\chi}_{\gamma\delta}(\tau-\tau')
  c_{\delta\beta}(\tau').
\end{eqnarray}
In the form of Eq.~(\ref{eq:socol2p}), the matrix
$\gamma_{\alpha\beta}(\tau)$ determines the master equation
(\ref{eq:seq}) and the reduced dynamics of the system. The matrix
$\gamma_{\alpha\beta}(\tau)$ may be represented by a sum of
Hermitian and anti-Hermitian matrices, {\em i.e.},
$\gamma_{\alpha\beta}(\tau) = \gamma_{\alpha\beta}^H(\tau) + i
\gamma_{\alpha\beta}^A(\tau)$. The Hermitian term
$\gamma_{\alpha\beta}^H(\tau)$ involves in the decoherence while
the anti-Hermitian term $i \gamma_{\alpha\beta}^A(\tau)$
contributes to the Hamiltonian dynamics of the system. We call the
Hermitian matrix $\gamma_{\alpha\beta}^H(\tau)$  a decoherence
rate matrix.  Now, the system operators in Eq.~(\ref{eq:socol2p})
are time-independent which enables the connection to the effective
environment approach.

The evolution superoperator ${\cal S}'(\tau)$, {\em i.e.}, the
solution to the master equation (\ref{eq:seq}), is not necessarily
completely positive {\em in the long-time limit}. On the other
hand, the method of the effective environment guarantees the
complete positivity of the evolution superoperator ${\cal
S}(\tau)$ {\em at all times}. In the Appendix we show how to
examine the complete positivity for a given superoperator. In the
effective environment approach one needs to determine the
parameters of the effective environment.  To determine them we
require that in the short-time limit the master equation derived
by the effective environment be approximately equal to the master
equation (\ref{eq:seq}) in conjuction with Eq.~(\ref{eq:socol2p})
. The effective environment approach may have different
higher-order terms from the time-convolutionless
projection-operator method and this difference leads to the
complete positivity of the reduced dynamics in the effective
environment approach.

Using the interaction Hamiltonian (\ref{eq:eeiih}) and following
the similar procedure leading to Eq.~(\ref{eq:seq}), we can derive
the master equation in the effective environment approach. This
has to be approximate to Eq.~(\ref{eq:seq}) in conjuction with
Eq.~(\ref{eq:socol2p}) in the short-time limit. This comparison
leads to the following two conditions
\begin{eqnarray}
  \label{eq:cond1}
  \mbox{Tr}_R \left[ \hat{D}_\alpha(\tau) \hat{\rho}_R(0) \right] = 0,
\end{eqnarray}
which is comparable to Eq.~(\ref{eq:ans}), and
\begin{eqnarray}
  \label{eq:cond2}
  \int^\tau_0 d\tau' \mbox{Tr}_R\left[{\hat D}_{\alpha}(\tau){\hat
  D}_{\beta}(\tau'){\hat \rho}_R(0)\right] \approx \gamma_{\alpha\beta}(\tau)
\end{eqnarray}
where $\hat{D}_\alpha(\tau)$ is given in Eq.~(\ref{eq:eeiih}). The
initial density operator $\hat{\rho}_R(0)$ and the time-dependent
coupling matrix $\lambda_{\alpha\beta}(\tau)$ can be chosen by varying
the operators $\hat{R}_\alpha$ so as to satisfy the coupled linear
equations (\ref{eq:cond1}) and (\ref{eq:cond2}). The Hilbert space of
the effective environment needs to be enlarged with respect to the
dimensionality unless any solution is found to Eqs.~(\ref{eq:cond1}) and
(\ref{eq:cond2}). As a result, the chosen set of operators
$\{\hat{D}_\alpha(\tau)\}$ are responsible for the type of decoherence
which the system undergoes. In this sense the operator
$\hat{D}_\alpha(\tau)$ is called a decoherence channel to the
environment.

In most cases, an effective environment has the same Hilbert-space
dimensionality as the quantum system. For instance, this is the case for
Markovian environments. Nonetheless, an effective environment can be of
larger dimensionality in order to properly describe the reduced dynamics
of a system. An effective environment may be classified into a
``$Q$-qubit environment'' if the effective environment consists of $Q$
qubits, ``$M$-mode bosonic environment'' if it consists of $M$ bosonic
modes, and so on.

We shall consider the asymptotic form of the time-correlation
function $\chi(\tau) = \frac{d}{d\tau}\gamma(\tau)$. The
time-correlation function is in general complex. However, being
interested in the decoherence, we concentrate on the real part of
the time-correlation function. The time-correlation function
becomes real-valued when the system Hamiltonian is redefined so
that it includes Lamb shifts \cite{Lamb}.  Noting that the
time-correlation disappears at a long time, $\chi(\tau)$
approaches zero in the limit of $\tau \rightarrow \infty$.  Thus,
$\chi(\tau)$ is assumed to have the form of
\begin{eqnarray}
  \label{eq:fotc}
  \chi(\tau) = f(\tau) e^{-g(\tau)}
\end{eqnarray}
where $f(\tau)$ and $g(\tau)$ are polynomial (or trigonometric)
functions and further $g(\tau)\rightarrow \infty$ as $\tau\rightarrow
\pm \infty$. A simple form of $\chi(\tau)$ will be given by
\begin{eqnarray}
  \label{eq:msfotc}
  \chi(\tau) = \frac{\kappa}{4 \tau_r} e^{-|\tau|/\tau_r},
\end{eqnarray}
where $\kappa = 1/\tau_d$ is a decoherence rate with the
decoherence time $\tau_d$ and $\tau_r$ is a relaxation time (or
memory time) in which the injected information is diffused over
the environment. In the Markovian limit of $\tau_r \ll \tau_d$,
the time correlation function becomes a delta function so that
$\chi(\tau) \rightarrow \kappa \delta(\tau)/2$ and $\gamma(\tau)
\rightarrow \kappa/4$.  The time-correlation function in
Eq.~(\ref{eq:msfotc}) may be regarded as the first extension from
Markovian to non-Markovian  decoherence \cite{Ahn02}.

\section{qubit system}

\label{sec:qubit}

We shall present typical decoherence of a qubit system in Markovian and
non-Markovian environments in the approach of an effective environment.
In addition, we investigate the structure of an effective environment:
a) the environmental variables that play a role in the respective
decoherence, b) the coupling constants between the system and the
effective environment, and c) the initial quantum states of the
effective environment.

We shall first of all consider a dephasing process. As in
Sec.~\ref{sec:dee}, we compare both master equations in the
projection-operator method and the effective environment approach for a
dephasing process. It is found that the effective environment has one
qubit and the interaction Hamiltonian is given by
\begin{eqnarray}
  \label{eq:qdpee}
  \hat{H}_{int}(\tau) &=& \lambda(\tau) \hat{\sigma}_z \otimes
  \hat{\sigma}_z, \\
  \label{eq:tdcc}
  \lambda(\tau)&=&\frac{2\gamma(\tau) \exp[-4\Gamma(\tau)]}
  {\sqrt{1-\exp[-8\Gamma(\tau)]}}
\end{eqnarray}
where $\hat{\sigma}_\alpha$ is a Pauli spin operator and
$\Gamma(\tau)=\int_0^\tau d\tau' \gamma(\tau')$. The initial states of
the system and the one-qubit effective environment can be written as
\begin{eqnarray}
  \label{eq:idos}
  \hat{\rho}_S(0) &=& \frac{1}{2} \left(\hat{\openone} + {\mathbf
  s}\cdot\hat{\boldsymbol{\sigma}} \right), \\
  \label{eq:idor}
  \hat{\rho}_R(0) &=& \frac{1}{2} \left(\hat{\openone} + {\mathbf
  r}\cdot\hat{\boldsymbol{\sigma}} \right),
\end{eqnarray}
where $\hat{\openone}$ is an identity operator, $\mathbf{s}=(s_x, s_y,
s_z)$ a Bloch vector of the system qubit, $\mathbf{r}=(r_x, r_y, r_z)$ a
Bloch vector of the one-qubit environment, and
$\hat{\boldsymbol{\sigma}} = (\hat{\sigma}_x, \hat{\sigma}_y,
\hat{\sigma}_z)$. In connection with the time correlation function of
the environment, the coupled linear equations (\ref{eq:cond1}) and
(\ref{eq:cond2}) are satisfied for $\mathbf{r} = (r_x,r_y,0)$ and
$\lambda(\tau)$ in Eq.~(\ref{eq:tdcc}), in the short-time limit of
$|\gamma(\tau)| \tau \ll 1$. The initial state of the effective
environment is unpolarized along the $z$ axis while it may be polarized
on the $x$-$y$ plane for the dephasing process.

\begin{figure}[htbp]
  \centering
  \includegraphics[width=0.45\textwidth]{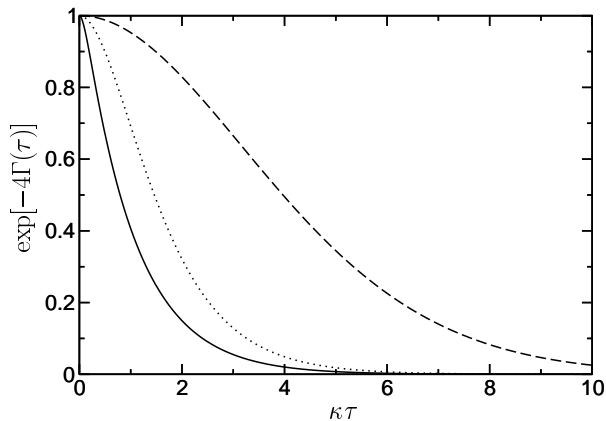}
  \caption{Decay function, $\exp[-4\Gamma(\tau)]$, with respect to the
    evolving time $\tau$. The decay function may be regarded as a fringe
    visibility for quantum interference as the time passes. It depends
    on the memory time $\tau_r$ for a given decoherence rate
    $\kappa=1/\tau_d$: a) $\kappa \tau_r=0.1$ (solid line), b) $\kappa
    \tau_r=1$ (dotted line), and c) $\kappa \tau_r=10$ (dashed line).
    The case a) is already close to the Markovian decoherence
    ($\kappa\tau_r \rightarrow 0$)}
  \label{fig:fig1}
\end{figure}

Let us consider the reduced dynamics of the system. Applying
Eqs.~(\ref{eq:qdpee}) and (\ref{eq:tdcc}) to
Eqs.~(\ref{eq:fseo})$-$(\ref{eq:rdos}), we obtain the reduced dynamics
in terms of its Bloch vector,
\begin{eqnarray}
  \label{eq:drdos}
  {\mathbf s}(\tau) = \left(s_x \exp[-4\Gamma(\tau)], s_y
  \exp[-4\Gamma(\tau)], s_z\right).
\end{eqnarray}
The dynamics of the dephasing process is characterized by the
decay function $\exp[-4\Gamma(\tau)]$. The decay function may be
regarded as a fringe visibility on the quantum interference for a
given time $\tau$. In order to compare the effect of the
non-Markovian decoherence to the Markovian one, we employ the
simple time-correlation function in Eq.~(\ref{eq:msfotc}). The
decay function then becomes
\begin{eqnarray}
  \label{eq:nmdec}
  \exp[-4\Gamma(\tau)] = \exp\left\{-\kappa\left[\tau -
  \tau_r \left(1-e^{-\tau/\tau_r} \right)\right] \right\}.
\end{eqnarray}
In the limit of no memory, $\tau_r \rightarrow 0$, the dephasing process
is Markovian with the decay function, $\exp[-4\Gamma(\tau)] =
\exp(-\kappa \tau)$.  When the environment keeps the information of the
system within the memory time $\tau_r$, the loss of the coherent
information is suppressed in comparison to the Markovian case as present
in Fig.~\ref{fig:fig1}.  This fact can be clearly seen through the
decoherence rate,
\begin{eqnarray}
  \label{eq:drfsn}
  \gamma(\tau) = \frac{d}{d\tau} \Gamma(\tau) = \frac{1}{4}
  \kappa\left(1-e^{-\tau/\tau_r} \right)
\end{eqnarray}
for a given memory time $\tau_r$. The decoherence rate is $\kappa/4$ if
$\tau_r \rightarrow 0$ and it is reduced by increasing the memory time
$\tau_r$.

Similar analysis can be applied to depolarizing or amplitude-damping
processes for a qubit. For this purpose, the interaction Hamiltonian is
assumed to be
\begin{eqnarray}
  \label{eq:qdee}
  \hat{H}_{int}(\tau) = \lambda(\tau) \sum_{\alpha,\beta=x,y,z}
  g_{\alpha\beta} \hat{\sigma}_\alpha \otimes \hat{\sigma}_\beta
\end{eqnarray}
where $\lambda(\tau)$ is given by Eq.~(\ref{eq:tdcc}) and
$g_{\alpha\beta}$ is a real coupling matrix which determines the type of
decoherence. The initial states of the system and the one-qubit
effective environment are given by their Bloch vectors ${\mathbf s}$ and
${\mathbf r}$, similar to Eqs.~(\ref{eq:idos}) and (\ref{eq:idor}).  The
coupling matrix $g_{\alpha\beta}$ and the initial state ${\mathbf r}$ of
the effective environment are summarized in Table.~\ref{tab:tab1}. It is
interesting to look at the initial state of the effective environment
for each process. The initial state is completely random with
$\mathbf{r}=(0,0,0)$ for the depolarizing process and it is polarized
along $z$ axis for the amplitude-damping process.

The reduce dynamics of the qubit is described in terms of the time
dependence of its Bloch vector ${\mathbf s}(\tau)$.  For the
depolarizing process, the time-dependent Bloch vector is given by
\begin{eqnarray}
  \label{eq:rdfdpol}
  {\mathbf s}_{pol}(\tau) =  {\mathbf s} \exp[- 8 \Gamma(\tau)]
\end{eqnarray}
For the amplitude damping process, it is given by
\begin{eqnarray}
  \label{eq:rdfad}
  \left[{\mathbf s}_{amp}(\tau)\right]_{x} &=& s_{x} \exp[-4\Gamma(\tau)]
  \nonumber \\
  \left[{\mathbf s}_{amp}(\tau)\right]_{y} &=& s_{y} \exp[-4\Gamma(\tau)]
  \nonumber \\
  \left[{\mathbf s}_{amp}(\tau)\right]_{z} &=& r_z + (s_z-r_z)
  \exp[-8\Gamma(\tau)].
\end{eqnarray}

It is well known that a system slowly decoheres in a non-Markovian
environment due to the memory effects. However, we stress that the
result should be derived from the completely positive reduced dynamics
and our analysis of the effective Hamiltonian approach strongly supports
the result.

\begin{table}
\caption{The coupling matrix $g_{\alpha\beta}$ and the initial
  state ${\mathbf r}$ of the effective environment for
  dephasing, depolarizing, and amplitude damping processes.}
\begin{ruledtabular}
\begin{tabular}{lcc}
 & $g_{\alpha\beta}$ & $\mathbf r$ \\
\colrule
dephasing & $\delta_{\alpha z} \delta_{\beta z}$ & $(r_x,r_y,0)$\\
depolarizing & $\delta_{\alpha \beta}$ & $(0,0,0)$ \\
amplitude damping & $\delta_{\alpha \beta}(\delta_{\beta x}+\delta_{\beta
  y})$ & $(0,0,r_z)$ \\
\end{tabular}
\label{tab:tab1}
\end{ruledtabular}
\end{table}

\section{Remarks}

The effective Hamiltonian approach was proposed to investigate the
non-Markovian decoherence of an open system and further to understand
the characteristics of its environment. The formalism is based on the
notion of an effective environment, {\em i.e.}, certain collective
degrees of freedom in the environment that are responsible for the
decoherence. The present approach naturally imposes the complete
positivity on the reduced dynamics for the system. We applied the
approach to the dephasing, depolarizing, and amplitude-damping
processes. It was found that the non-Markovian environment suppresses
the decoherence of the qubit, due to the memory effect of the
non-Markovian environment.

\acknowledgments

This work was supported by the UK Engineering and Physical Science
Research Council and by the Korean Ministry of Science and Technology
through the Creative Research Initiatives Program under Contract
No. M10116000008-03F0000-03610. J.L. acknowledges the support from the
KOSEF through Quantum Photonic Science Research Center and the research
fund by Hanyang University (HY-2003). H.McA. thanks Department of
Education and Learning for support.

\appendix

\section{Complete positivity of a superoperator}
\label{sec:vsbo}

Consider a set of bounded operators acting on the state vector in the
Hilbert space. The set of bounded operators forms a vector space $V$
with the inner product defined by Hilbert-Schmidt norm \cite{Reed80}.
The inner product of two bounded operators $\hat{v}$ and $\hat{w}$ in
$V$ is given by
\begin{eqnarray}
  \label{eq:ipbo}
  \llv{\hat{w}}|\rrv{\hat{v}} \equiv \mbox{Tr}\left(\hat{w}^\dagger
  \hat{v} \right).
\end{eqnarray}
In the analogy of the ket and bra states, $|\rrv{\hat{v}}$ is called a
``ket'' vector of an operator $\hat{v}$ and $\llv{v}|$ a ``bra'' vector.
The set of the bra vectors forms the dual vector space.  Let
$\{\hat{e}_{ab}=|\rv{a}\lv{b}|\}$ be an orthonormal basis for the vector
space with the completeness relation,
\begin{eqnarray}
  \label{eq:crbov}
  \sum_{ab} |\rrv{\hat{e}_{ab}}\llv{\hat{e}_{ab}}| = \so{I}
\end{eqnarray}
where $\so{I}$ is an identity superoperator, {\em i.e.}, a linear map of
the ket vector (or ket operator) to itself.  Any ket vector
$|\rrv{\hat{v}}$ is expanded as
\begin{eqnarray}
  |\rrv{\hat{v}} = \sum_{ab} v_{ab}|\rrv{\hat{e}_{ab}}
\end{eqnarray}
where $v_{ab} = \llv{\hat{e}_{ab}}|\rrv{\hat{v}}$.

A superoperator $\so{S}$ is a linear map of $V$ onto itself
\begin{eqnarray}
  \so{S} : \hat{v} \rightarrow \hat{v}' =
  \so{S}(\hat{v}).
\end{eqnarray}
The matrix elements of $\so{S}$ in the orthonormal basis
$\{\hat{e}_{ab}\}$ is obtained as
\begin{eqnarray}
  \label{eq:meso}
  S_{ab,cd} = \llv{\hat{e}_{ab}}|\so{S}|\rrv{\hat{e}_{cd}} =
  \llv{\hat{e}_{ab}}|\rrv{\so{S}(\hat{e}_{cd})}.
\end{eqnarray}
The superoperator $\so{S}$ is called ``Hermitian'' when
\begin{eqnarray}
  \label{eq:hso}
  S_{ab,cd} = S^*_{cd,ab}.
\end{eqnarray}
A superoperator $\so{S}$ is called ``positive'' when
$\llv{\hat{\eta}}|\so{S}|\rrv{\hat{\eta}}$ is real and positive for all
$\hat{\eta} \in V$.

If $\so{S}$ is Hermitian, it has the right eigenvector
$|\rrv{\hat{v}_s}$ with the ``real'' eigenvalue $\lambda_s$ such that
\begin{eqnarray}
  \label{eq:revso}
  \so{S} |\rrv{\hat{v}_s} = \lambda_s |\rrv{\hat{v}_s}.
\end{eqnarray}
and $\llv{\hat{v}_s}|$ is the corresponding left eigenvector with
the same eigenvalue. Further, if and only if $\so{S}$ is positive,
the eigenvalues are all positive. When $\so{S}$ is not Hermitian,
it may have complex eigenvalues and $\llv{\hat{v}_s}|$ is no
longer the corresponding left eigenvector. Instead, there exists a
left eigenvector $\llv{\hat{w}_s}|$ such that
\begin{eqnarray}
  \label{eq:levso}
  \llv{\hat{w}_s}| \so{S} = \llv{\hat{w}_s}| \lambda_s.
\end{eqnarray}
where $\lambda_s$ is a complex number. The right and left eigenvectors satisfy
the orthogonality
\begin{eqnarray}
  \label{eq:og}
  \llv{\hat{w}_s}|\rrv{\hat{v}_{s'}} = \llv{\hat{w}_s}|\rrv{\hat{v}_s}
  \delta_{ss'}.
\end{eqnarray}
The completeness relation is given as the following
\begin{eqnarray}
  \label{eq:csrli}
  \sum_{s} \frac{|\rrv{\hat{v}_s} \llv{\hat{w}_s}|}
  {\llv{\hat{w}_s}|\rrv{\hat{v}_s}} = \so{I}
\end{eqnarray}
and $\so{S}$ can be represented as
\begin{eqnarray}
  \label{eq:rlerso}
  \so{S} = \sum_{s} \lambda_s \frac{|\rrv{\hat{v}_s} \llv{\hat{w}_s}|}
  {\llv{\hat{w}_s}|\rrv{\hat{v}_s}}.
\end{eqnarray}

{\em Definition}: Partial transposition, denoted by ``\#'', on a
superoperator $\so{S}$ is defined as
\begin{eqnarray}
  \label{eq:ptos}
  {S}^\#_{ab,cd} = \llv{\hat{e}_{ab}}| \so{S}^\# |\rrv{\hat{e}_{cd}} =
  \llv{\hat{e}_{ac}}| \so{S} |\rrv{\hat{e}_{bd}} = S_{ac,bd}.
\end{eqnarray}
A superoperator $\so{S}$ is said to be Hermitian-preserving when
$\so{S}(\hat{x}^\dagger) = \so{S}(\hat{x})^\dagger$ for any $\hat{x} \in
V$. The partial transposed superoperator $\so{S}^\#$ is Hermitian if and
only if $\so{S}$ is Hermitian-preserving. In addition, $\so{S}^\#$ is
closely related to the complete positivity of $\so{S}$.

{\em Theorem}: The following three conditions for a superoperator
$\so{S}$ are all equivalent:
\begin{itemize}
\item[(a)] $\so{S}$ is completely positive
\item[(b)] $\so{S}$ has a Kraus representation, {\em i.e.},\\
  \begin{eqnarray}
    \so{S}(\hat{x}) = \sum_\mu \hat{K}_\mu \hat{x} \hat{K}^\dagger_\mu
  \end{eqnarray}
\item[(c)] $\so{S}^\#$ is positive
\end{itemize}
{\em Proof}: It is well known \cite{Kraus71} that (a) and (b) are
equivalent. Consider (b) $\Rightarrow$ (c). When $\so{S}$ has a
Kraus representation, its matrix elements are given as
\begin{eqnarray}
  S_{ab,cd} = \sum_\mu K^{ac}_\mu {K^{bd}_\mu}^*.
\end{eqnarray}
The partial transposition $\so{S}^\#$ becomes
\begin{eqnarray}
  S^\#_{ab,cd} = S_{ac,bd} = \sum_\mu K^{ab}_\mu {K^{cd}_\mu}^*.
\end{eqnarray}
It is clear that for all $\hat{\eta} \in V$
\begin{eqnarray}
  \llv{\hat{\eta}}|\so{S}^\#|\rrv{\hat{\eta}}= \sum_{abcd} \eta^*_{ab}
  S^\#_{ab,cd} \eta_{cd} = \sum_\mu X_\mu X_\mu^* > 0
\end{eqnarray}
where $X_\mu=\sum_{ab} \eta^*_{ab} K^{ab}_\mu$.  Thus $\so{S}^\#$ is
positive.

The converse (c) $\Rightarrow$ (b) is now considered. Since $\so{S}^\#$
is positive, it is also Hermitian and all the eigenvalues $d_\nu$ are
positive:
\begin{eqnarray}
  S^\#_{ab,cd} = \sum_\nu d_\nu
  \llv{\hat{e}_{ab}}|\rrv{\hat{v}_\nu}\llv{\hat{v}_\nu}|\rrv{\hat{e}_{cd}}
\end{eqnarray}
where $|\rrv{\hat{v}_\nu}$ are normalized eigenvectors.  Since $d_\nu$
is positive, we can define the following matrix
\begin{eqnarray}
  \tilde{K}^{ab}_\nu = \sqrt{d_\nu} \llv{\hat{e}_{ab}}|\rrv{\hat{v}_\nu}.
\end{eqnarray}
Thus the superoperator $\so{S}$ has a Kraus representation as
\begin{eqnarray}
  S_{ab,cd} = S^\#_{ac,bd} = \sum_\nu (\tilde{K}^{ac}_\nu)
  ({\tilde{K}^{bd}_\nu})^*.
\end{eqnarray}

The set of Kraus operators are not unique \cite{Kraus71} and all
equivalent sets of Kraus operators for the given superoperator can be
generated by ``unitary remixing'' of the canonical set with the
eigenvalue vector ${\bf d}'$ extended by some arbitrary number of zeros,
{\em i.e.} ${\bf d}'=({\bf d},0,...,0)$ \cite{Hughston93}.


\begin{thebibliography}{99}

\bibitem{Jozsa94} R. Jozsa, J. Mod. Opt. {\bf 41}, 2315 (1994); A.
  Peres, \pra {\bf 61}, 022116 (2000); J. H. Reina, L. Quiroga and N.
  F. Johnson, \pra {\bf 65}, 032326 (2002).

\bibitem{Schumacher96} B. Schumacher, \pra {\bf 54}, 2614 (1996).

\bibitem{Lidar98} D. A. Lidar, I. L. Chuang, and K. B. Whaley, \prl {\bf
    81}, 2594 (1998).

\bibitem{Lee00} J. Lee and M. S. Kim, \prl {\bf 84}, 4236 (2000); M.  S.
  Kim and J. Lee, \pra {\bf 64}, 012309 (2001).

\bibitem{Zurek03} W. H. Zurek, \rmp {\bf 75}, 715 (2003).

\bibitem{Cirac96} J. F. Poyatos, J. I. Cirac, and P. Zoller, \prl {\bf
    77}, 4728 (1996); S. Bose, P. L. Knight, M. B. Plenio, and V.
  Vedral, \prl {\bf 83}, 5158 (1999).

\bibitem{Myatt00} C. J. Myatt, B. E. King, Q. A. Turchette,
  C. A. Sackett, D. Kielpinski, W. M. Itano, C. Monroe, and
  D. J. Wineland, Nature {\bf 403}, 269 (2000).

\bibitem{Gardiner92} C. W. Gardiner, {\em Quantum Noise}
  (Springer-Verlag, Berlin, 1992); R. Alicki and K. Lendi, {\em Quantum
    Dynamical Semigroups and Applications} (Springer-Verlag, Berlin,
  1987).

\bibitem{Kane98} B. E. Kane, Nature {\bf 393}, 133 (1998); N. Vats and
  T.  Rudolph, \jmo {\bf 48}, 1495 (2001); W. D.  Oliver, F. Yamaguchi,
  and Y. Yamamoto, \prl {\bf 88}, 037901 (2002).

\bibitem{John97} T. Quang, M. Woldeyohannes, S. John, and G. S.
  Agarwal, \prl {\bf 79}, 5238 (1997); S. John and T. Quang \prl {\bf
    78}, 1888 (1997).

\bibitem{Hope00} J. J. Hope, G. M. Moy, M. J. Collect, and C. M.
  Savage, \pra {\bf 61}, 023603 (2000).

\bibitem{Ahn97} D. Ahn, Prog. Quantum Electron {\bf 21}, 249 (1997) and
  references therein.
  
\bibitem{Knezevic03} I. Knezevic and D. K. Ferry, \pre {\bf 67}, 066122
  (2003); I. Knezevic and D. K. Ferry, \pre {\bf 66}, 016131 (2002).

\bibitem{Ahn02} D. Ahn, J. Lee, M. S. Kim, and S. W. Hwang, \pra {\bf
    66}, 012302 (2002).

\bibitem{Lee02} M. S. Kim, J. Lee, D. Ahn, and P. Knight, \pra {\bf 65},
  040101(R) (2002).

\bibitem{Braun02}, D. Braun, \prl {\bf 89}, 277901 (2002).

\bibitem{Stinespring55} W. F. Stinespring, Proc. Amer. Math. Soc. {\bf
    6}, 211 (1955).

\bibitem{Kraus71} K. Kraus, Ann. Phys. {\bf 64}, 311 (1971).

\bibitem{Pechukas94} P. Pechukas, \prl {\bf 73}, 1060 (1994); R.
  Alicki, \prl {\bf 75}, 3020 (1995); P. Pechukas, \prl {\bf 75}, 3021
  (1995).

\bibitem{Lindblad76} G. Lindblad, Commun. Math. Phys. {\bf 48}, 119
  (1976).

\bibitem{Bacon99} D. Bacon, D. A. Lidar, and K. B. Whaley, \pra {\bf
    60}, 1944 (1999).

\bibitem{Daffer03} S. Daffer, K. W\'odkiewicz, J. D. Cresser, and
  J. K. McIver, e-print quant-ph/0309081.

\bibitem{McAneney03} H. McAneney, J. Lee, and M. S. Kim, \pra {\bf 68},
  063814 (2003); H. Fearn and R. Loudon, Opt. Commun. {\bf 64}, 485
  (1987); M. S. Kim and N. Imoto, \pra {\bf 52}, 2401 (1995).

\bibitem{Ahn94} D. Ahn, \prb {\bf 50}, 8310 (1994).

\bibitem{Ahn00} D. Ahn, J. H. Oh, K. Kimm, and S. W. Hwang, \pra {\bf
    61}, 052310 (2000).

\bibitem{Caldeira85} A. O. Caldeira and A. J. Leggett, \pra {\bf 31},
  1059 (1985).

\bibitem{Zwanzig60} R. Zwanzig, J. Chem. Phys. {\bf 33}, 1338 (1960); M.
  Saeki, Prog. Theor. Phys. {\bf 67}, 1313 (1982).

\bibitem{Lamb} Here, the Lamb shift is undistinguished from the Stark
  shift, although the Lamb shift results from the vacuum fluctuation
  whilst Stark shift is from the thermal fluctuation (see
  Ref.~\cite{Gardiner92}).

\bibitem{Barnett97} S. M. Barnett and P. M. Radmore, {\em Methods in
    Theoretical Quantum Optics} (Clarendon Press, Oxford, 1997).

\bibitem{Reed80} M. Reed and B. Simon, {\em Methods of Modern
    Mathematical Physics} (Academic, London, 1980), Vol. 1.

\bibitem{Hughston93} L. P. Hughston, R. Jozsa, and W. K. Wootters, Phys.
  Lett. A {\bf 183}, 14 (1993).

\end{thebibliography}
\end{document}